\documentclass[preprint2]{aastex}

\usepackage[dviout]{graphics/graphicx}
\usepackage{longtable}
\usepackage{lscape}
\usepackage{multirow}
\usepackage{color}
\usepackage{ulem}

\shorttitle{The origin of the near-infrared excess in SN Ia 2012dn: Circumstellar dust around the super-Chandrasekhar supernova candidate}
\shortauthors{Nagao et al.}

\begin{document}

\title{The origin of the near-infrared excess in SN Ia 2012dn: Circumstellar dust around the super-Chandrasekhar supernova candidate}

\author{Takashi Nagao\altaffilmark{1,4}, Keiichi Maeda\altaffilmark{1,2}, Masayuki Yamanaka\altaffilmark{3}
}
\altaffiltext{1}{Department of astronomy, Kyoto University, Kitashirakawa-Oiwake-cho,
 Sakyo-ku, Kyoto 606-8502, Japan}
\altaffiltext{2}{Kavli Institute for the Physics and Mathematics of the 
Universe (WPI), The University of Tokyo, 
5-1-5 Kashiwanoha, Kashiwa, Chiba 277-8583, Japan}
\altaffiltext{3}{Department of Physics, Faculty of Science and Engineering, Konan University, 8-9-1 Okamoto, Kobe, Hyogo 658-8501, Japan}
\altaffiltext{4}{Email: nagao@kusastro.kyoto-u.ac.jp}

\begin{abstract}
  The nature of progenitors of the so-called super-Chandrasekhar candidate Type Ia supernovae (SC-SNe Ia) has been actively debated. Recently, Yamanaka et al. (2016) reported a near-infrared (NIR) excess for SN 2012dn, and proposed that the excess originates from an echo by circumstellar (CS) dust. In this paper, we examine a detailed distribution of the CS dust around SN 2012dn, and investigate implications of the CS dust echo scenario for general cases of SC-SNe Ia. We find that a disk/bipolar CS medium configuration reproduces the NIR excess fairly well, where the radial density distribution is given by a stationary mass loss. The inner radius of the CS dust is $0.04$ pc. The mass-loss rate of the progenitor system is estimated to be $1.2 \times 10^{-5}$ and $3.2 \times 10^{-6}$ M$_{\odot}$ yr$^{-1}$ for the disk and bipolar CS medium configurations, respectively, which adds another support for the single degenerate scenario. Our models limit SN 2009dc, another SC-SN Ia, to have a dust mass less than $0.16$ times that of SN 2012dn. While this may merely indicate some variation on the CS environment among SC-SNe Ia, this could raise another interesting possibility. There could be two classes among SC-SNe Ia; the brighter SC-SNe Ia in a clean environment (SN 2009dc) and the fainter SC-SNe Ia in a dusty environment (SN 2012dn).
\end{abstract} 


\section{Introduction}
	Type Ia supernovae (SNe Ia) have been used as a powerful standardizable candle to measure cosmological distances (Riess et al. 1998; Perlmutter et al. 1999). This relies on the homogeneous properties of SNe Ia as is characterized by a relation between the decline rate of the light curve and the absolute magnitude (Phillips 1993), which is believed to originate from the homogeneous nature of the progenitors, presumably being nearly Chandrasekhar-mass white dwarfs (WDs). 

	 The progenitor evolution leading to SNe Ia is, however, still obscured (see, e.g., Maeda \& Terada 2016 for a review). Circumstellar media (CSM) around SNe Ia provide important diagnostics to reveal the nature of their progenitors. The progenitor systems should have relatively dense CSM in the so-called single-degenerate (SD) scenario, where the companion star is a non-degenerate star (Whelan \& Iben 1973; Nomoto 1982). The progenitor systems in the SD scenario are expected to eject a large amount of materials through various processes (e.g. mass loss from the companion stars, nova explosions, binary interaction). On the other hand, the progenitor systems in the double-degenerate (DD) scenario (a pair of two WDs; Iben \& Tutukov 1984; Webbink 1984) would not create CSM environment as dense as expected in the SD scenario. If the progenitor systems have dense CSM as predicted in the SD scenario, such CSM could be detectable through an echo, ether a thermal near-infrared (NIR) emission (Maeda et al. 2015b) or an optical scattering echo by dust grains in the CSM. For normal SNe Ia, such signatures have not been detected so far (Maeda et al. 2015b).

         However, there is an emerging diversity of SNe Ia, and it is possible that SNe Ia come from multiple progenitor populations. In this sense, it is highly interesting to test the existence of CSM around various sub-types of SNe Ia. In this paper, we focus on a class of SNe called `super-Chandrasekhar' candidate SNe Ia (SC-SNe Ia, or `overluminous' SNe Ia), that do not follow the relation between the peak luminosity and the declining rate. This class of SNe Ia includes the following objects: SNe 2003fg, 2006gz, 2007if, and 2009dc (Howell et al. 2006; Hicken et al. 2007; Yamanaka et al. 2009; Tanaka et al. 2010; Yuan et al. 2010; Scalzo et al. 2010; Silverman et al. 2011; Taubenberger et al. 2011; Scalzo et al. 2012). They show different properties from normal SNe Ia, typically showing high luminosity at the light-curve peak, slow evolution in their light curves at early phases, low expansion velocities and strong carbon features (see Yamanaka et al. 2009 for the well-observed and prototypical SC-SN Ia 2009dc). They also show different behaviors in late phases from normal SNe Ia, characterized by low ionization state and temperature (Maeda et al. 2009; Taubenberger et al. 2013). Recently, some SNe, showing the similar properties with the prototypical SC-SNe except for the brightness, have been discovered (e.g. SNe 2011aa, 2012dn and possibly 2006gz; Brown et al. 2014; Chakradhari et al. 2014). The luminosity of these SNe is comparable with that of normal SNe Ia. However, they are believed to be in the same class as the brighter SC-SNe, given their striking similarities in the other spectral and light-curve properties.

	The nature of progenitors of the SC-SNe Ia has been actively debated. Typically more than one solar mass of $^{56}$Co (created as $^{56}$Ni) is needed to explain their peak luminosity, assuming that their energy source is the $^{56}$Co decay. Therefore, the ejecta and progenitor masses are suggested to exceed the Chandrasekhar mass (e.g. Howell et al. 2006; Yamanaka et al. 2009). They are thus termed super-Chandrasekhar SNe Ia, which might be possible for rapidly rotating massive WD progenitors (Taubenberger et al. 2011; Hachinger et al. 2012; Kamiya et al. 2012). We note that there are other proposed scenarios to explain their large luminosity by introducing alternative energy sources, such as interaction between SN ejecta and CSM (e.g. Hachinger et al. 2012; Noebauer et al. 2016) or release of the thermal energy stored in an extended envelope of a WD as a result of a WD-WD merger (Scalzo et al. 2010, 2012), or considering a viewing angle effect on an off-center explosion of a Chandrasekhar-mass WD explosion (Hillebrandt et al. 2007).

	 As for the CSM around a progenitor of SC-SNe Ia, an NIR excess was recently discovered for SN 2012dn (Yamanaka et al. 2016). SN 2012dn is classified as a SC-SN Ia, while its luminosity is comparable to normal SNe Ia (see above; Bock et al. 2012; Chakradhari et al. 2014; Yamanaka et al. 2016). Yamanaka et al. (2016) proposed that the NIR excess is originated from absorption and reemission of the SN light by circumstellar (CS) dust. As a demonstration, they compared the observed NIR excess with a theoretical prediction by Maeda et al. (2015b), where they considered thermal emission from CS dust distributed within an infinitely thin shell. The distance to the CS dust shell is limited to be $4.8-6.4 \times 10^{-2}$ pc, and the mass of the CS dust shell is estimated to be $10^{-6}-10^{-5}$ M$_{\odot}$ yr$^{-1}$. However, the predicted light curve evolution from their simple spherical model shows significant difference to the observed NIR light curves (see \S2). Furthermore, they did not consider the scattering process by CS dust. Thus, there are two issues which are still open; (1) If we can obtain detailed information on the distribution of CS dust, and (2) If this explanation of the NIR excess is consistent with the fact that there was no additional luminosity in the optical bands. In general, if an SN has CS dust in the vicinity of the SN, an additional optical luminosity due to the scattering echo by the CS dust should be generated (e.g. Wang 2005; Goobar 2008; Nagao et al. 2016). 
        
	In this paper, we examine the detailed distribution of the CS dust around SN 2012dn, and also investigate whether the optical light curves of SN 2012dn are consistent with the expected scattering echo from the CS dust. In \S 2, we summarize observational properties of SN 2012dn. In \S 3, we describe our updated echo model for CS dust, which includes both thermal and scattering echoes from CS dust. In \S 4.1, the results for SN 2012dn are presented. Based on the results for SN 2012dn, we further investigate general behaviors that the CS dust echo model predicts. In \S 4.2, we discuss effects of a viewing angle on the NIR and optical excesses. In \S 4.3, we examine implications for a prototypical SC-SN Ia 2009dc. The paper concludes in \S 5 with discussion.

\section{Observational Properties of SC-SN Ia 2012dn}
        One aim of this paper is to investigate details of the CS dust echo scenario for an NIR excess observed for SC-SN Ia 2012dn. The observational properties of SN 2012dn are described in Yamanaka et al. (2016), and in this section we summarize main features of this object. The absolute luminosity of SN 2012dn in the optical bands are lower than those of SN 2009dc, which is a well-observed and prototypical SC-SN Ia. However, the optical light curves of SN 2012dn are indeed quite similar with those of SN 2009dc, if the light curves of SN 2009dc are dimmed by $\sim 0.5$ magnitude. This suggests that the intrinsic light curves of SN 2012dn are similar to those of SN 2009dc with the difference only in the absolute luminosity.

          For the NIR light curves, the situation is the same until $\sim 30$ days from the $B$-band maximum, being nearly identical to those of SN 2009dc dimmed by $\sim 0.5$ magnitude. Thereafter, the NIR light curves of SN 2012dn show deviation from this behavior toward higher luminosity. (see Yamanaka et al. 2016). Yamanaka et al. (2016) produced the `scaled' light curves of SN 2009dc, by dimming its early-phase light curve in each NIR band pass to match to that of SN 2012dn in the same band pass. This scaled light curves are assumed to represent the intrinsic SN component of SN 2012dn. 
      
          These behaviors indicate that the intrinsic light curves of SNe 2012dn and 2009dc are similar (except for the luminosity), but an additional component is associated with the late-time NIR emission from SN 2012dn. This is further supported by the color evolution, which is free from the difference in the luminosity. The $V-J$, $V-H$ and $V-K_s$ color evolutions of SN 2012dn are almost identical with those of SN 2009dc until $\sim 30$ days after the $B$-band maximum, and thereafter SN 2012dn becomes redder (Fig. 4 in Yamanaka et al. 2016). The same behavior is observed for the NIR-NIR color curves. From these behaviors, Yamanaka et al. (2016) concluded that the intrinsic NIR light curves of SN 2012dn are similar to the scaled light curves of SN 2009dc, and attributed the difference of the late-phase light curves in the NIR bands to an additional component, i.e., the NIR excess. Note that the NIR excess requires an additional component in the spectral energy distribution (SED), since a single blackbody spectrum cannot explain the NIR luminosity after $\sim 30$ days from the $B$-band maximum (e.g., Fig. 9 of Yamanaka et al. 2016). 

\begin{figure}[t]
\includegraphics[width=9.0cm]{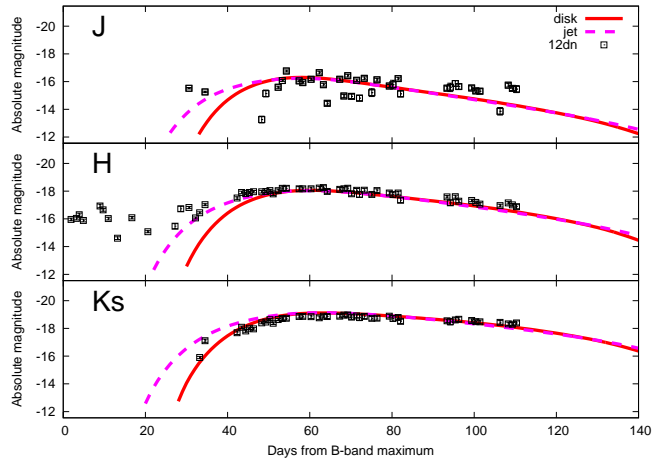}
\caption{The $J$-, $H$- and $K_{s}$-band light curves of the NIR excess component of SN 2012dn (black square points). Our best-fit echo models are shown for the disk case (red solid lines) and for the bipolar jet case (magenta dotted lines).}
\end{figure}
          
	In Figure 1, we show the light curves of the NIR excess seen in SN 2012dn, which are derived by subtracting the scaled light curves of SN 2009dc from the light curves of SN 2012dn (Yamanaka et al. 2016). We note that the subtracted light curves (i.e., the excess light curves) shown in Fig. 1 have noisy structures especially in the early phase. This mainly comes from the fact that two SNe are very similar in the early phase without clear excess, and therefore the subtraction essentially leads to zero flux. On the other hand, the late-phase data (after $\sim 30$ days from the $B$-band maximum) are dominated by the excessive component (especially in the $H$ and $K_{s}$ bands), and details of the template for the subtraction (i.e., the scaled light curves of SN 2009dc) do not sensitively affect the derived NIR excess component. Namely, the derived NIR excess light curves are quite robust and insensitive to the assumed intrinsic component. 

	Yamanaka et al. (2016) attributed the NIR excess to a thermal echo from CS dust grains, on the basis of a rough agreement between the observed excess and the echo prediction, in the light-curve shape of the excess (a flat shape) and the corresponding temperature of the excess ($1200-2200$ K). They placed a limit on the CS dust mass and the distance from the progenitor, using a spherically distributed thin shell model by Maeda et al. (2015b). However, a spherical distribution should always lead to the emergence of the echo before the $B$-band maximum, and therefore the model was unable to explain the delayed emergence of the NIR excess at $\sim 30$ days after the $B$-band maximum. In addition, a thin shell model should always lead to an exactly flat light curve, while the observed excess shows gradual decay in its flux. These discrepancies would be remedied by introducing a more detailed configuration than the spherical thin shell, and indeed this provides a great opportunity to investigate the detailed configuration of CSM around SNe Ia which is linked to the progenitor evolution scenarios.

\section{Model}
	In this section, we describe our dust echo model, which is an updated version of the model by Maeda et al. (2015b). In this updated model, thermal and scattering echoes from CS dust are self-consistently calculated for an arbitrary spatial distribution of the CS dust. In calculating the echoes, we need information on the incident SN light, optical properties of the CS dust and the distribution/amount of the CS dust. Once these parameters are given, we can calculate a light curve of an echo from the CS dust, for which our formalism is given in \S 3.3. In this paper, we fix the optical properties of the CS dust and the incident SN light. 

The optical properties of the CS dust used in this paper are described in \S 3.1. For the incident SN light, we adopt the Hsiao's template spectral evolution as the input SN light (Hsiao et al. 2007). The Hsiao's template spectral series, as constructed by a number of normal SNe Ia, covers the spectral evolution from early to late phases in the UV to NIR band passes, some of which are missing in observational of individual SNe. Therefore, if a target SN (i.e., SN 2012dn in this paper) has a reasonably similar spectral evolution to the Hsiao's template, it is tempting to use this as the incident SN light. In fact, the light curves of SN 2012dn in the optical bands are quite similar with those in the Hsiao's template, despite the `spectral' identification of SN 2012dn as a SC-SN Ia. 

A non-negligible difference between the Hsiao's template and the observed light curves of SN 2012dn is found in the UV and NIR. However, as the incident SN light used in the echo simulation, the difference in these band passes is not important. While the UV flux is larger in SN 2012dn than in the Hsiao's template (or normal SNe Ia: Brown et al. 2014), the total flux in the UV is still quite lower than that in the optical band, so it would not affect the thermal balance of the CS dust. In the NIR bands, the opacity is low (\S 3.2) and therefore the absorption of the intrinsic NIR light provides at most a minor contribution to the total energy. We note that, when the combination of the intrinsic SN and the NIR echo is required in our analysis, the echo contribution is added to the scaled light curves of SN 2009dc (not to the Hsiao's template) after the echo emission is created, while the incident SN light is replaced by the Hsiao's template.

With these parameters fixed, the only remaining parameters are for the distribution and amount of the CS dust (\S 3.2). In this paper, we vary these as free parameters and simulate the light curves of thermal and scattering echoes. By comparing the resulting light curves to those observed for SN 2012dn, we aim to derive the distribution and amount of CS dust.

\subsection{Dust model}

\begin{table}[t]
  \caption{Dust parameters in our dust model}
  \small
\begin{tabular}{ccccc}
\tableline
 filter & $\lambda$[$\mu$ m] & $\kappa_{\mathrm{abs}}$[cm$^{2}$/g] & $\kappa_{\mathrm{scat}}$[cm$^{2}$/g] & g\\
\tableline \tableline
U & 0.36 & 5.539E+4 & 3.518E+4 & 4.038E-1\\ \hline
B & 0.44 & 4.601E+4 & 3.095E+4 & 3.592E-1\\ \hline
V & 0.55 & 3.706E+4 & 2.625E+4 & 3.203E-1\\ \hline
R & 0.66 & 3.014E+4 & 2.153E+4 & 3.115E-1\\ \hline
I & 0.81 & 2.269E+4 & 1.617E+4 & 2.678E-1\\ \hline
J & 1.2  & 1.318E+4 & 8.317E+3 & 1.725E-1\\ \hline
H & 1.7  & 8.202E+3 & 3.431E+3 & 5.452E-2\\ \hline
Ks & 2.2 & 5.283E+3 & 1.243E+3 & 2.887E-2\\ 
\tableline
\end{tabular}
\end{table}

For the CS dust model, we adopt graphite grains, because as Yamanaka et al. (2016) pointed out that, assuming that the NIR excess in SN 2012dn is created by radiation from the dust, the derived dust temperature is too high for non-carbonaceous dust (e.g. Nozawa \& Kozasa 2013). In our dust model, the shape of the dust grains is assumed to be spherical. The size distribution follows the MRN size distribution, $\propto a^{-3.5}$, where $a$ is a radius of a dust grain (Mathis et al. 1977). The maximum and minimum radii of the dust grains are 0.2 $\mu$m and 0.05 $\mu$m, respectively. Using these values, the values of the scattering opacity ($\kappa_{\mathrm{scat},\nu}$), the absorptive opacity ($\kappa_{\mathrm{abs},\nu}$) and the scattering angle distribution ($g$) in the dust model are uniquely determined through Mie theory. In Table 1, the parameters used for the dust are summarized.

 In fact, the dust composition and the grain size distribution for CS dust around SNe Ia are not clear, while adopting the graphite composition is observationally robust (see above). To simplify the problem under consideration, in our dust model as described above, the parameters related to the dust size distribution are set to be typical values frequently used in the literature. A good agreement between the observed NIR colors and the predicted echo colors suggests that the adopted wavelength dependence of the absorptive opacity as a function of the wavelength, as determined by the size distribution,  is fairly well represented by our model (see \S 4). Also, a different size distribution would not dramatically change the total brightness of the thermal and scattering echoes, since the scale of the opacity at the SN peak wavelength is mostly determined by the dust composition and insensitive to the size distribution as long as the typical size of the dust is smaller than the wavelength of interest. A relatively large uncertainty would be only involved in the ratio of the brightness between the thermal and scattering echoes, since a different size distribution could change the albedo in the blue portion of the optical wavelength (see \S 5 for further discussion).

 \subsection{Distribution of CS dust}
\begin{figure*}[t]
  \includegraphics[width=15.0cm]{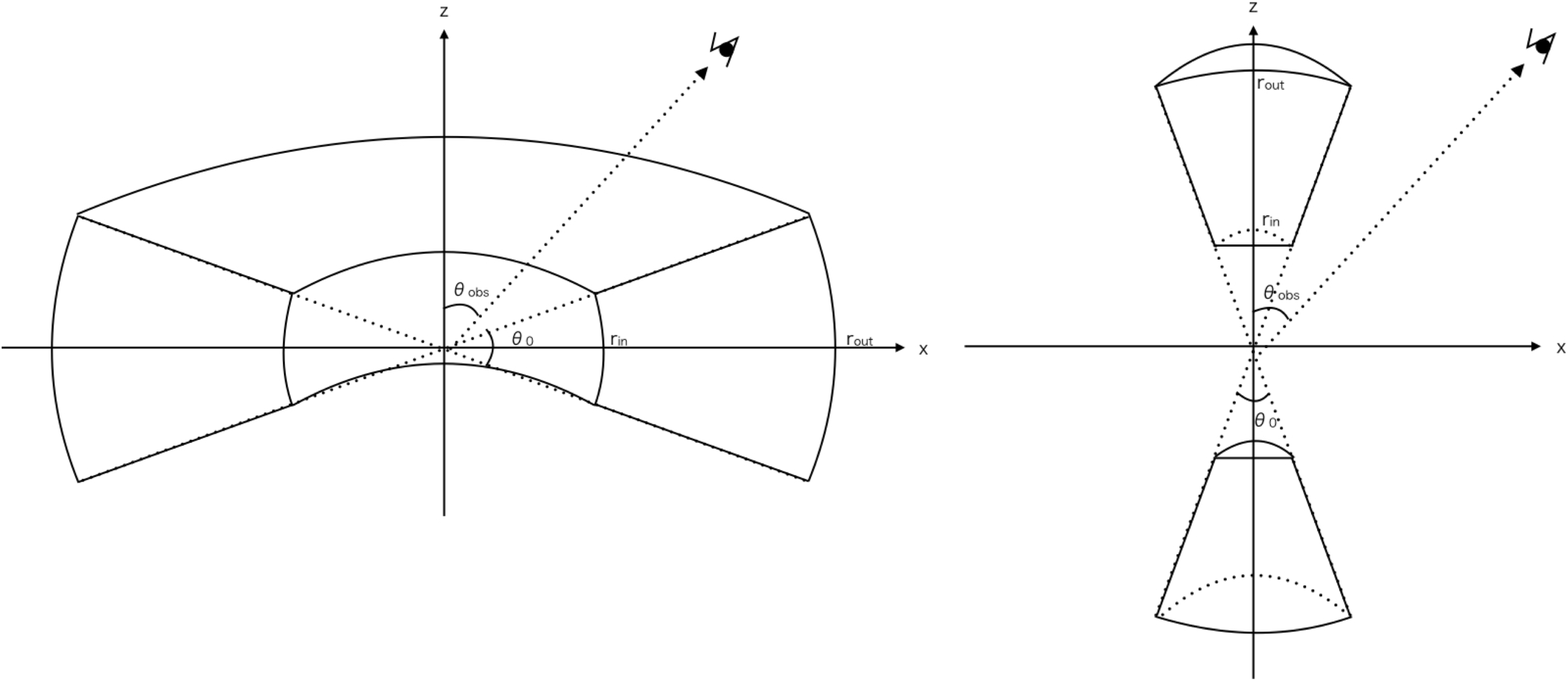}
  \vspace{-1cm}
\caption{The geometry of the CS dust distribution adopted in this study; the disk (left) and the bipolar jet (right).}
\end{figure*}
 
	Given the shortcoming of a spherically distributed CS dust to explain the delayed emergence of the NIR excess as observed for SN 2012dn (\S 2; see also Maeda et al. 2015b; Yamanaka et al. 2016 for details), we consider two configurations which are still simple but motivated by the progenitor scenarios; the disk and bipolar jet models (see Fig. 2), as inferred for some progenitor systems in the SD scenario (e.g. Dilday et al. 2012; Chesneau et al. 2012). In the disk model, the inner and outer radii are denoted by  $r_{\mathrm{in}}$ and $r_{\mathrm{out}}$, respectively, and the opening angle of the disk is denoted by $\theta_{0}$. Here, $r_{\mathrm{in}}$ should be larger than the evaporation radius of graphite grains determined by the SN peak luminosity ($\sim 0.02$ pc for our graphite dust model). Even if the mass loss of a progenitor continues until just before an SN explosion, the CS dust below the evaporation radius should be destroyed and make no contribution to the echo, and thus $r_{\mathrm{in}}$ in this case corresponds to the evaporation radius. In our model, we vary $r_{\mathrm{in}}$ simply as a free parameter, and its consistency to the evaporation radius is checked/discussed afterword. The observer$^{\prime}$s direction is expressed as an angle $\theta_{\mathrm{obs}}$, which is an angle between the observer$^{\prime}$s direction and the polar direction of the disk. In the bipolar jet model, the inner and outer radii of the jet are expressed as $r_{\mathrm{in}}$ and $r_{\mathrm{out}}$, respectively, and the opening angle of the jet is expressed as $\theta_{0}$. The observer$^{\prime}$s direction is expressed as an angle $\theta_{\mathrm{obs}}$, which is an angle between the observer$^{\prime}$s direction and the polar direction of the jet. 

In both models, the radial density distribution of the CS dust is assumed to follow $\rho_{\mathrm{dust}}(r) = \rho_{\mathrm{dust}}(r_{\mathrm{in}}) (r/r_{\mathrm{in}})^{-2}$, as expected from a stationary mass loss from a progenitor or a companion star. This might also give a good approximation for episodic mass loss (e.g. novae) as averaged over a time scale of thousands years, which is a scale probed by an echo.

\subsection{Thermal and scattering echoes from CS dust}

	In calculating thermal and scattering echoes from CS dust, we divide the CS dust into small fragments on the 3D grids. The echoes from each fragment are calculated separately, and then each contribution is integrated for the observer's frame. We first describe the formalism in the optically thin limit, and then introduce correction factors to take into account main effects of the extinction within the CSM.
	
	First, we consider a situation where dust grains absorb SN photons that are originally emitted at time $t$ (the time since the explosion in the observer's frame) and emit the energy as thermal emission. Assuming radiative equilibrium, the temperature $T(t,r)$ of the dust grains at position ($r,\theta,\phi$), is given by        
\begin{equation}
\int^{\infty}_{0} \frac{L_{\mathrm{SN},\nu}(t)}{4\pi r^{2}}\kappa_{abs,\nu}d\nu = 4\pi \int^{\infty}_{0} \kappa_{abs,\nu}B_{\nu}(T(t,r))d\nu ,
\end{equation}
where $L_{\mathrm{SN},\nu}(t)$ is an incoming SN luminosity that are emitted from the SN at time $t$, $B_{\nu}(T)$ is the Plank function with temperature $T$, and $\kappa_{abs,\nu}$ is a mass absorption coefficient of the dust grains at frequency $\nu$. Note that the temperature is not dependent on $(\theta, \phi)$, reflecting our assumption of the optically thin limit for the incoming SN light (see below for more details). Furthermore, we note that in the present model the temperature is dependent on the radial coordinate, unlike the model by Maeda et al. (2015b) who considered an infinitely thin shell.

Once the temperature is given, the luminosity of the thermal emission from the small fragment at the position $(r,\theta,\phi)$ is given by
\begin{equation}
  4\pi \kappa_{abs,\nu} B_{\nu}(T(t,r)) \rho_{\mathrm{dust}} (r,\theta,\phi) r^{2} dr \sin \theta d\theta d\phi.
\end{equation}
This emission reaches to an observer with time delay with respect to the SN photons that are emitted at the same time $t$ but reaching to the observer directly without further interactions. The time delay, $\Delta t$, is given as follows:
\begin{equation}
  \Delta t = \frac{r(1-\cos \theta^{\prime})}{c},
\end{equation}
where $\theta^{\prime}$ is the angle between the observer direction and the position vector to the fragment $(r,\theta,\phi)$.

	Next, we consider the scattering echo in the same situation as the thermal emission. For dust grains at a position ($r,\theta,\phi$) to scatter SN photons emitted at time $t$, the luminosity of the scattering echo at the fragment is given as follows:
\begin{equation}
 L_{SN,\nu}(t) \times \frac{d\Omega}{4\pi}(\kappa_{scat,\nu}\rho_{\mathrm{dust}}(r,\theta,\phi)dr) \times p(\theta^{\prime}) \times 4 \pi,
\end{equation}
where $p(\theta^{\prime}) \times 4 \pi$ is a term to take into account a scattering angle distribution. Here, $p(\theta)$ is set by the Henyey-Greenstein scattering angle distribution (Henyey \& Greenstein 1941) as follows:
\begin{eqnarray}
p(\theta) = \frac{1}{4\pi} \frac{1-g^{2}}{(1+g^{2}-2g\cos \theta)^{\frac{3}{2}}}, \; \; \; \; \; \mathrm{and}\\
\int_{0}^{2\pi} \int_{0}^{\pi} p(\theta) \sin \theta d\theta d\phi = 1 \ , 
\end{eqnarray}
where $g$ is uniquely determined for a given dust model (\S 3.1). 

 The above formulations are derived for the optically thin limit. 
 For dense CS dust, the optical depth may not be negligible. There are two processes to consider as the optical depth effect: 
\begin{enumerate}
\item Extinction of the incident SN light by CS dust.
\item Extinction of the thermal and scattering echoes by CS dust
\end{enumerate}
The first process affects the flux of incident SN light, as an input to the thermal and scattering echoes, at different positions. The second effect is to alter the emergent echo spectra, especially a relative strength between the thermal and scattering echoes. In principle, these effects can be self-consistently taken into account by performing a full radiation transfer. However, in the model used in this paper, we calculate contributions from each fragment to the total echo emissions separately, which are integrated with the time delay effect in the end of the simulation. Performing a full radiation transfer, which connects the transfer between different fragments, is beyond a scope of this paper. 

However, including the first-order correction to these effects is still possible within our formulations. By neglecting a secondary diffuse light and only considering the extinction of primary photon rays (both the SN light and echo), this is done by adding an additional exponential term, $exp(-\tau_\nu)$ to the left-hand-side (L.H.S) of equation 1 (for the incident SN light), to formula 2 (for the thermal echo) and formula 4 (for the scattering echo). 

For the extinction of the incident SN light by CS dust, we decided not to include this correction term for the sake of simplicity. The main reason is the following. This effect should work on the optical and NIR echoes in the same way and would not alter the relative strength, as the incident flux to create the thermal/scattering echoes is both proportional to $exp(-\tau_\nu)$ where the wavelengths of interest are both in the optical wavelength. This effect would therefore alter only the normalization of the derived mass loss rate (which is inversely proportional to the incident flux): Indeed, this would lead to underestimate of the mass loss rate, at most by a factor of two. 

On the other hand, we include the optical depth effect to the echo emission as follows. The optical depth between a fragment ($r, \theta, \phi$) and an observer ($\theta_{\mathrm{obs}}, \phi_{\mathrm{obs}}$) is given as follows: 
\begin{equation}
\tau_{\mathrm{obs},\nu} (r,\theta,\phi,\theta_{\mathrm{obs}},\phi_{\mathrm{obs}}) = \int_{0}^{\infty} \kappa_{\mathrm{ext},\nu} \rho_{\mathrm{dust}}(s) ds \ ,
\end{equation}
where $s$ is a distance from the fragment ($r, \theta, \phi$) along the observer's direction. The optical depth here is dependent on the position of the fragment and the viewing direction, but further simplification is possible. The largest effect arises at the innermost position within the dusty CS environment due to the high density and the strongest contribution to the echo emissions, and indeed the optical depth at the mid-plane of the disk/jet in the inner radius provides a good approximation to that at any point in the CS disk/jet (within a factor of a few). We therefore adopt this value as a characteristic optical depth ($\tau_{\mathrm{ch},\nu}$) in taking into account the optical depth effect. This characteristic optical depth, $\tau_{\mathrm{ch},\nu}$, can be described as $\tau_{\mathrm{obs},\nu} (r_{\mathrm{in}},\pi/2,0,0,0)$. In our calculations, the NIR and optical echo luminosity (i.e., formulae 2 and 4) are extinguished by the characteristic optical depth, and after that the contributions from each fragment are integrated by taking into account the time delay. This optical depth effect works on these echoes in a different manner, because the characteristic optical depth is proportional to the opacity, which is sensitive to the frequency. Namely, the optical scattering echo should suffer from much larger extinction than the NIR thermal echo. While our treatment is simple, this includes the main effect of the extinction and allows us to robustly discuss the difference between the NIR and optical echoes.

\section{Results}

\subsection{SN 2012dn}

\begin{table*}[t]
  \caption{Best-fit parameters}
\begin{tabular}{ccccccc}
\tableline
 model & $\theta_{0}$ & $r_{\mathrm{in}}$[pc] & $r_{\mathrm{out}}$[pc] & $\theta_{\mathrm{obs}}$ & $\rho_{\mathrm{dust}}(r_{\mathrm{in}})$ [g/cm$^{3}$] & $\dot{M}_{\mathrm{gas}}$[M$_{\odot}$/yr]\\
\tableline \tableline
disk & 20 & 0.04 & 0.1 & 0 & 2.4E-22 & 1.2E-5\\ \hline
jet & 20 & 0.04 & 0.1 & 90 & 7.1E-22 & 3.2E-6\\ 
\tableline
\end{tabular}
\end{table*}

The best-fit parameters for the NIR excess of SN 2012dn are determined by comparing the observed NIR light curves and the simulated thermal echo following the formalisms in \S 3, where the model free parameters are $\theta_0, r_{\mathrm{in}},r_{\mathrm{out}} , \theta_{\mathrm{obs}}$ and $\rho_{\mathrm{dust}}(r_{\mathrm{in}})$. Figure 1 shows the NIR excess components in the light curves of SN 2012dn (Yamanaka et al. 2016), compared with our best-fit disk and bipolar jet models. The NIR excess of SN 2012dn is constructed by subtracting the scaled light curves of SN 2009dc from those of SN 2012dn (\S 2; see Yamanaka et al. 2016 for details). The best-fit parameters are shown in Table 2.

     The delayed emergence of the NIR echo limits the configuration strongly. The line of sight must be nearly perpendicular to the bulk of the CS dust distribution, i.e., either the pole-on view for the disk distribution or the side view for the bipolar distribution. As such, the CS dust does not contribute the extinction of SN 2012dn.

The derived CS dust distribution can be converted to the (gas) mass loss rate for a given dust-to-gas ratio ($f_{\rm dust}$) and the progenitor wind velocity ($v_{\mathrm{wind}}$) as follows:
\begin{eqnarray}
  \dot{M}_{\mathrm{gas}} (r) &=& \biggl( \frac{\rho_{\mathrm{dust}}(r)}{f_{\rm dust}} \biggr) \biggl( 2\pi \int d\theta^{'} \sin \theta^{'} \biggr) r^2 v_{\mathrm{wind}} \nonumber \\
  &=& \frac{\rho_{\mathrm{dust}} (r_{\mathrm{in}}) r_{\mathrm{in}}^{2}}{f_{\rm dust}} \biggl( 2\pi \int d\theta^{'} \sin \theta^{'} \biggr) v_{\mathrm{wind}},
\end{eqnarray}
where the integration is performed for an appropriate range in $\theta^{'}$ (see Fig. 1). Because we assume $\rho_{\mathrm{dust}} (r) \propto r^{-2}$ corresponding to a stationary mass loss (which indeed provides a good fit to the observed NIR light curves; see below), $\dot{M}_{\mathrm{gas}} (r)$ does not depend on $r$. In Table 2, we assume $f_{\rm dust} = 0.01$ and $v_{\mathrm{wind}} = 10$ km s$^{-1}$. 

\begin{figure}[t]
\includegraphics[width=9cm]{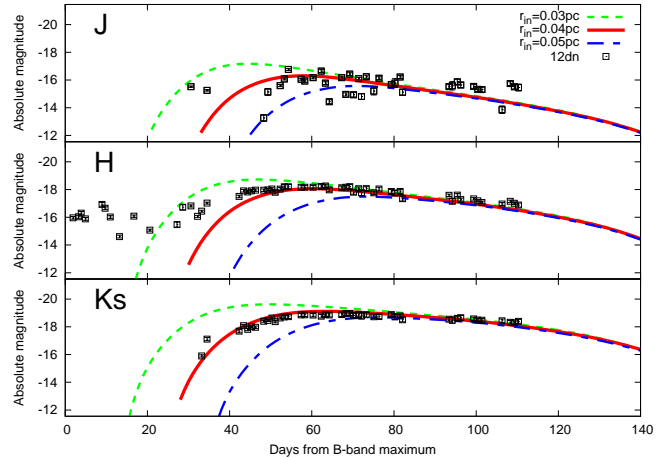}
\caption{The $J$-, $H$- and $K_{s}$-band light curves of the NIR echo for the disk model, with various values of $r_{\mathrm{in}}$ (red solid lines for $r_{\mathrm{in}} = 0.04$ pc, green dotted lines for $r_{\mathrm{in}} = 0.03$ pc, blue dash-dotted lines for $r_{\mathrm{in}} = 0.05$ pc). The light curves of the NIR excess seen in SN 2012dn are also shown (black square points).}
\end{figure}
           
	Figure 3 shows the light curves of the NIR echo for the disk model with various values of $r_{\mathrm{in}}$. This figure shows that the value of $r_{\mathrm{in}}$ strongly affects the rising part of the echo light curve, and the inner radius, $r_{\mathrm{in}}$, can be accurately determined to be $0.04$ pc.

\begin{figure}[t]
\includegraphics[width=9cm]{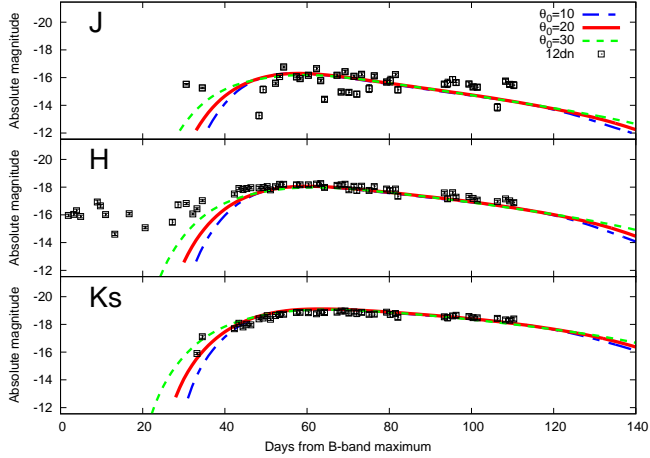}
\caption{The $J$-, $H$- and $K_{s}$-band light curves of the NIR echo for the disk model, with various values of $\theta_{0}$ (red solid lines for $\theta_{0} = 20$ degrees, green dotted lines for $\theta_{0} = 30$ degrees, blue dash-dotted lines for $\theta_{0} = 10$ degrees). The black square points are the same as in Fig. 3.}
\end{figure}

	Figure 4 shows the light curves of the NIR echo for the disk model with various values of $\theta_{0}$. For $\theta_{0} = 10 \sim 30$, it does not strongly affect the light curve of the NIR echo. The value of  $r_{\mathrm{out}}$ either makes an insignificant change of the light curve of the NIR echo until 100 days, if the value of $r_{\mathrm{out}}$ is bigger than $0.1$ pc. The minimum value of $r_{\mathrm{out}}$ is $0.1$ pc to explain the NIR echo in SN 2012dn. The values of $\theta_{0}$ and $r_{\mathrm{out}}$ cannot be further limited by the currently available light curves of SN 2012dn. In other words, the interpretation of the NIR excess in SN 2012dn as the echo by CS dust is insensitive to these details.

\begin{figure}[t]
  \includegraphics[width=9cm]{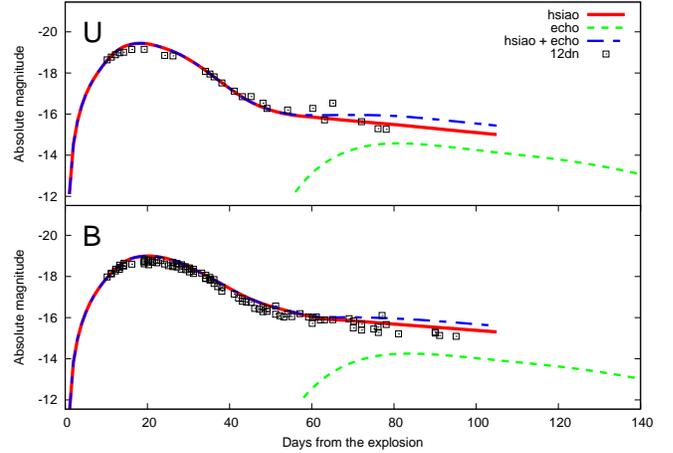}
\caption{The $U$- and $B$-band light curves predicted by the best-fit disk model for the NIR light curves of SN 2012dn. Shown here are the Hsiao's SN template (red solid lines), the echo contributions (green dotted lines) and the sum of the SN and echo contributions (blue dash-dotted lines). The light curves for SN 2012dn are shown by black square points.}
\end{figure}

	There is no clear excess in the optical light curves of SN 2012dn (Chakradhari et al. 2014, Yamanaka et al. 2016). In order to check the consistency between the NIR and optical light curves of SN 2012dn within the echo scenario, we calculate the optical scattering echo in the disk model with the best-fit parameters. Figure 5 shows the $U$- and $B$-band light curves of the scattering echo in the best-fit disk model. The optical scattering echo is hidden by the intrinsic SN light, and therefore the echo scenario is consistent with the observations of SN 2012dn both in the NIR and optical properties.

        \subsection{Echoes from the CS dust at different viewing angles}
\begin{figure}[t]
\includegraphics[width=9cm]{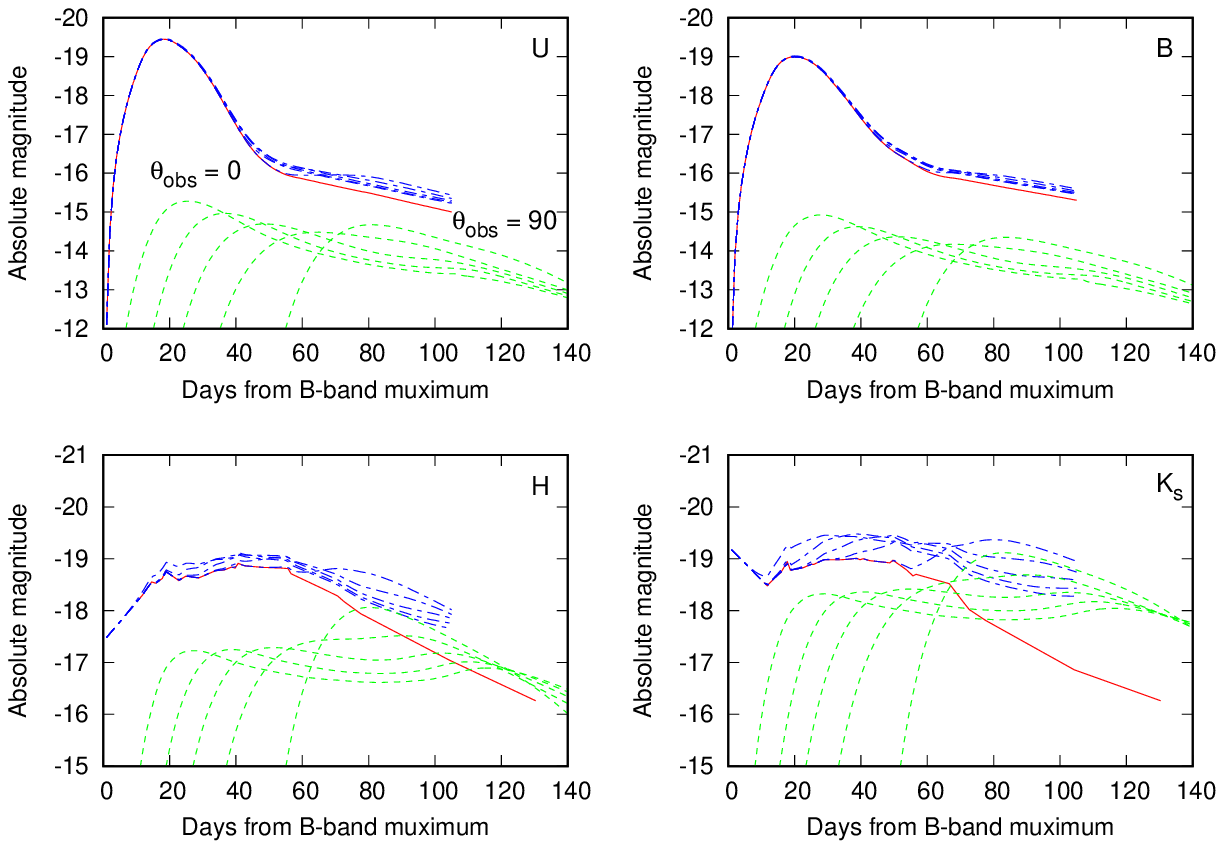}
\caption{The $U$-, $B$-, $H$- and $K_{s}$-band light curves based on the best-fit disk model for SN 2012dn but with various values of $\theta_{\mathrm{obs}}$ ($\theta_{\mathrm{obs}} = 0, 30, 45, 60, 90$). For the intrinsic SN light (red solid lines), we adopt the light curves of the Hsiao's template in the optical ($U$ and $B$ bands) and the light curves of SN 2009dc in the NIR ($H$ and $K_{s}$ bands). The light curves of the echoes are shown in green dotted lines. The blue dash-dotted lines show the sum of the SN and echo contributions.}
\end{figure}

\begin{figure}[t]
\includegraphics[width=9cm]{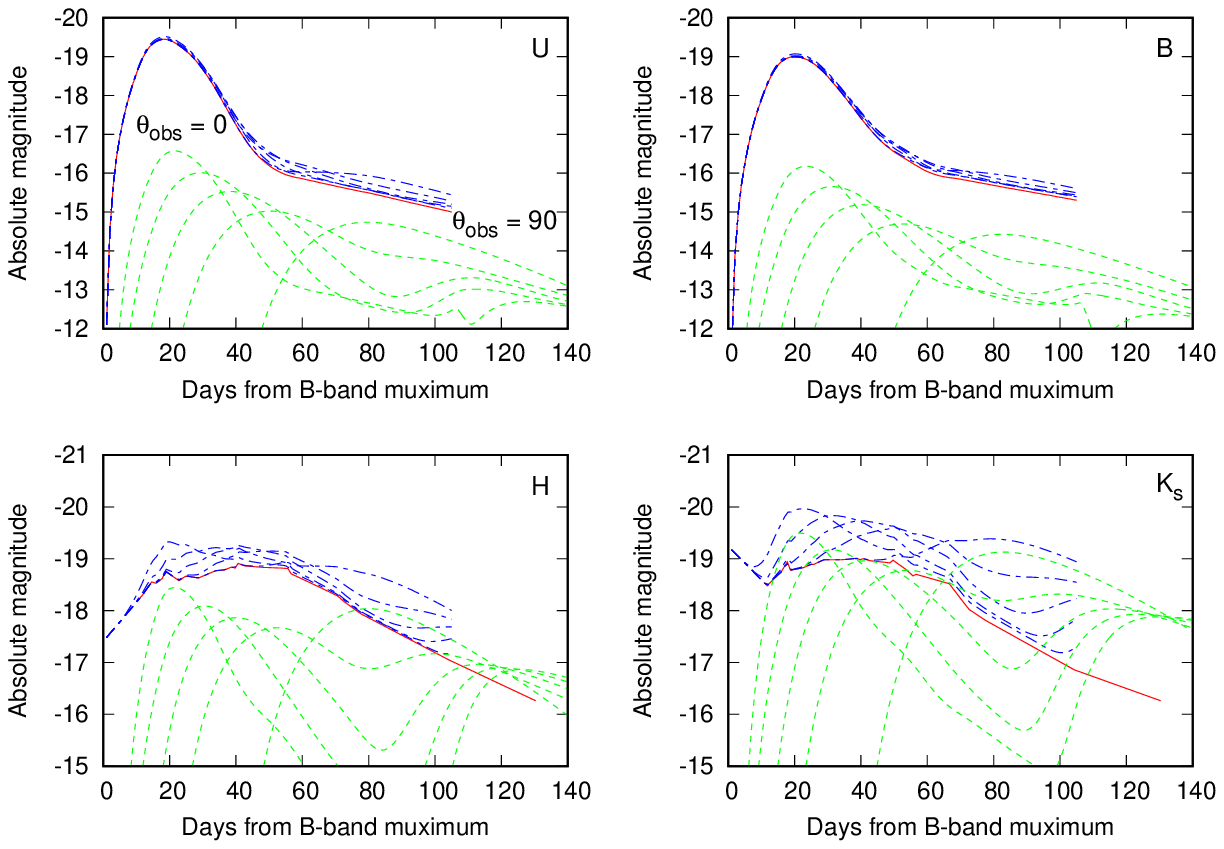}
\caption{Same as Fig. 6, but for the best-fit bipolar jet model.}
\end{figure}

	If the dusty CS disk (or jet) found for SN 2012dn is an universal feature in SC-SNe Ia, such a system will be observed from various viewing directions, once a large sample of SC-SNe Ia is constructed. Therefore, we calculate the thermal and scattering echoes from the CS dust with various viewing angle ($\theta_{\mathrm{obs}}$), fixing the other parameters to the best-fit model for SN 2012dn (Fig. 6 and Fig. 7 for the disk and bipolar jet models, respectively). If viewed from a different direction, the NIR light curves are predicted to be different from those in SN 2012dn. On the other hand, there is no model where we can observe clear optical excess. If there is a system where the amount of CS dust is substantially larger than that of SN 2012dn, such a system would also create detectable optical echoes. However, we emphasize that there is no model that shows excess only in the optical bands, as long as we adopt the same dust properties as SN 2012dn. 

	An interesting situation is the following; If viewed from a direction such that the CS disk/jet is between an SN and an observer ($80 \leq \theta_{\mathrm{obs}} \leq 100$), the intrinsic SN light (mostly emitted in the optical) is also extinguished by the CS disk/jet. In such a case, the contribution of the echo light relative to the SN light becomes more significant in the optical than in the NIR. However, in this case, the peaks of the scattering echo and the SN light curves must be overlapped ($\Delta t$ must be small), which means that observing the optical echo contribution is practically difficult also in this case. 

\subsection{SN 2009dc}
	In this study, we assume that the light curves of SN 2009dc are purely consist of the SN light without an echo, following Yamanaka et al. (2016) (see also \S 2). However, if SN 2009dc would also have dusty CS environment like SN 2012dn, there could already be some contribution of the echo in the NIR light curves of SN 2009dc. This will slightly increase the CS dust mass required to explain the NIR light curves of SN 2012dn. At the same time, an interesting question is how much CS dust might have existed around SN 2009dc.

	The most conservative upper limit of the mass of CS dust is obtained by the requirement that the predicted echo luminosity should not exceed the observed one at any bands and epochs. If SN 2009dc would have the identical CS disk to that around SN 2012dn, this limit is determined by the observed luminosity in the $K_s$ band at the late phase ($\gtrsim 130$ days since the explosion), where the relative luminosity of the echo to the SN is largest. To place the most conservative limit, we choose $\theta_{\mathrm{obs}} = 0$ for the comparison, as this situation leads to the lowest echo luminosity at $\gtrsim 130$ days. Taking into account the fact that the absolute luminosity of SN 2009dc in the optical and NIR bands are larger than that of SN 2012dn and typical SNe Ia, we estimate the maximally allowed mass of the CS disk for SN 2009dc. The estimated upper limit is $\sim 0.16$ times the CS dust mass derived for SN 2012dn. While it may merely indicate some variation on the CS environment among SC-SNe Ia, this could raise another interesting possibility. There could be two classes among SC-SNe Ia; the brighter SC-SNe Ia in a clean environment (SN 2009dc) and the fainter SC-SNe Ia in a dusty environment (SN 2012dn).

         With a large amount of CS dust around SN 2012dn, one may wonder if the difference in the luminosity of SN 2012dn and 2009dc would be attributed to the extinction through the CS dust - the answer is no. First, the CS environment derived for SN 2012dn does not contribute to the extinction of SN 2012dn, since the CS dust must not exist along the line of sight to explain the delayed emergence of the NIR excess (S 4.1). Moreover, this interpretation is not consistent with the fact that the difference in the luminosity of these SNe is the similar value ($\sim 0.5$ mag) in both the optical and NIR bands. Therefore, we conclude that the intrinsic luminosity of these SNe, with the different CS environments, is different from each other.
        
\section{Discussion and conclusions}
	In this paper, we have shown that the NIR excess observed in SN 2012dn can be naturally explained by an echo from CS dust around the SN. We have further derived the detailed distribution of the CS dust around SN 2012dn. In addition, we have confirmed that the optical observations of SN 2012dn are consistent with the CS echo model that explains the NIR excess.

	For a progenitor system of SNe Ia, there are two popular models; a binary of a WD and a non-degenerate star (the SD scenario) and a binary of two WDs (the DD scenario). For SN 2012dn, the SD scenario is favored from the large amount of CSM. We have derived the mass loss rate of $1.2 \times 10^{-5}$ M$_{\odot}$ yr$^{-1}$ for the disk model and $3.2 \times 10^{-6}$ M$_{\odot}$ yr$^{-1}$ for the bipolar jet model, assuming that the dust-to-gas ratio is 0.01 and the progenitor wind velocity is 10 km s$^{-1}$. Moreover, the estimated distribution of the CSM (either the disk or the bipolar jet) adds another support for the SD scenario. For example, the CSM have been suggested to be confined to the equatorial plane in an SN Ia showing the interaction between the SN ejecta and CSM (Dilday et al. 2012). Also, the bipolar CSM distribution is observed for some nova systems (Chesneau et al. 2012).

	In principle, emissions powered by a collision between the ejecta and the CSM in SN 2012dn may be detectable for a relatively dense CSM derived from the echo emission. If the SN luminosity determines the inner radius of the CS dusty disk by the evaporation of the CS dust, there may be gas below the inner radius derived in this paper. In fact, the derived inner radius of the CS disk ($\sim 0.04$ pc) is consistent with the evaporation radius of graphite grains by the SN peak luminosity ($\sim 0.02$ pc for our graphite dust model), given the uncertainties in the grain size distribution and the evaporation temperature (e.g. Dwek 1985). 

	We roughly estimate the expected magnitude due to the interaction, by scaling values of each physical parameters for SN 2012dn to those derived for an SN with clear signatures of the interaction. We use the values for Type IIb SN 2013df for normalizations, for which the derived CSM density (Maeda et al. 2015a) is at the same order to that derived for SN 2012dn. Note that this comparison has nothing to do with the origin of the CSM and progenitor systems which must be different between SNe Ia and IIb, but for the luminosity from the SN-CSM interaction the most important parameter is the CSM density irrespective of the origin of CSM. The luminosity due to the ejecta-CSM interaction is proportional to the fifth power of the ejecta velocity and the mass loss rate of the progenitor; $L \propto \dot{M} r_{\mathrm{shock}}^2 v_{\mathrm{shock}}^3 \sim \dot{M} v_{\mathrm{shock}}^5$ (e.g. Moriya et al. 2013). Assuming that $v_{\mathrm{shock}}(\mathrm{Ia})/v_{\mathrm{shock}}(\mathrm{IIb}) \propto \sqrt{E(\mathrm{Ia})/M(\mathrm{Ia})}/\sqrt{E(\mathrm{IIb})/M(\mathrm{IIb})} \sim \sqrt{2}$, we estimate that the luminosity of SN 2012dn due to the interaction might become larger than the SN itself after $\sim 400$ days since the explosion, and the $R$-band magnitude at that time might be $\sim 21$ mag in the disk model. For the jet configuration, it might be $\sim 22.5$ mag at $\sim 500$ days. On the other hand, if there is physically no material below the inner radius of the CS disk, the interaction will start when the ejecta reach the inner radius ($\sim 0.04$ pc) of the CS disk. Assuming that the ejecta velocity is 10000 km s$^{-1}$, the interaction would already have started around July 2016. The $R$-band magnitude at that time is estimated to be $\sim 24$ mag in the disk model and $\sim 25.5$ mag in the bipolar jet model. We may be able to observe these phenomena using $8$m-class telescopes like the Subaru telescope. Thus, if available, investigating the late time light curve of SN 2012dn is highly interesting. Furthermore, observing late-time light curves of future SC-SNe Ia is highly encouraged. 

	We have also calculated the expected echo light curves of SNe Ia that have exactly the same CS dust (the same composition, mass and distribution) as SN 2012dn except for the viewing angles. If the CS dust is common among SC-SNe Ia, we will find SNe that show the diverse properties in the NIR light curves as shown in Figures 6 and 7. We also conclude that there is no model that predicts an excess only in the optical bands irrespective of the mass and distribution of the CS dust, if we adopt the same dust properties used for SN 2012dn. Only with different properties of the CS dust grains, it is possible to realize a situation where only an optical echo would be detected. If such a detection would be found in the future for different SC-SNe Ia, it should indicate a diverse property of the CS dust grains. This may happen depending on the evaporation process of the dust grains or the creation history of the CS dust.

	We have also discussed the possibility that SN 2009dc might have dusty CS environment like SN 2012dn. We conclude that the maximally allowed mass of CS dust is $\sim 0.16$ times the mass for SN 2012dn. While it may merely indicate some variation on the CS environment among SC-SNe Ia, this could raise another interesting possibility. There could be two classes among SC-SNe Ia; the brighter SC-SNe Ia in a clean environment (SN 2009dc) and the fainter SC-SNe Ia in a dusty environment (SN 2012dn).

\acknowledgments
	The simulations were in part carried out on the PC cluster at Center for Computational Astrophysics, National Astronomical Observatory of Japan. The work has been supported by Japan Society for the Promotion of Science (JSPS) KAKENHI Grant 26800100 (K.M.). The work by K.M.\ is partly supported by World Premier International Research Center Initiative (WPI Initiative), MEXT, Japan. The work by M.Y. is partly supported by the Hirao Taro Foundation of the Konan University Association for Academic Research.


\end{document}